\documentclass[twocolumn,
amsmath,amssymb,prl,
aps,superscriptaddress,longbibliography
]{revtex4-2}

\usepackage{graphicx}
\usepackage{dcolumn}
\usepackage{bm}
\usepackage{braket}
\usepackage{mathtools}
\usepackage{bbm}
\usepackage{comment}
\usepackage{makecell}
\usepackage{afterpage}
\usepackage{siunitx}

\newcommand\blankpage{%
    \null
    \thispagestyle{empty}%
    \addtocounter{page}{-1}%
    \newpage}

\usepackage{hyperref}
\hypersetup{
	colorlinks=true,
	linkcolor=blue,
	citecolor=blue,
	filecolor=magenta,
	urlcolor=cyan
}

\def\iu{{\rm i}}

\DeclareMathOperator{\Tr}{Tr}

\def\dif{{\rm d}}

\graphicspath{{../}{tikz/}}

\makeatletter
\DeclareFontFamily{OMX}{MnSymbolE}{}
\DeclareSymbolFont{MnLargeSymbols}{OMX}{MnSymbolE}{m}{n}
\SetSymbolFont{MnLargeSymbols}{bold}{OMX}{MnSymbolE}{b}{n}
\DeclareFontShape{OMX}{MnSymbolE}{m}{n}{
	<-6>  MnSymbolE5
	<6-7>  MnSymbolE6
	<7-8>  MnSymbolE7
	<8-9>  MnSymbolE8
	<9-10> MnSymbolE9
	<10-12> MnSymbolE10
	<12->   MnSymbolE12
}{}
\DeclareFontShape{OMX}{MnSymbolE}{b}{n}{
	<-6>  MnSymbolE-Bold5
	<6-7>  MnSymbolE-Bold6
	<7-8>  MnSymbolE-Bold7
	<8-9>  MnSymbolE-Bold8
	<9-10> MnSymbolE-Bold9
	<10-12> MnSymbolE-Bold10
	<12->   MnSymbolE-Bold12
}{}

\let\llangle\@undefined
\let\rrangle\@undefined
\DeclareMathDelimiter{\llangle}{\mathopen}%
{MnLargeSymbols}{'164}{MnLargeSymbols}{'164}
\DeclareMathDelimiter{\rrangle}{\mathclose}%
{MnLargeSymbols}{'171}{MnLargeSymbols}{'171}
\makeatother

\def\ox{Rudolf Peierls Centre for Theoretical Physics, Clarendon Laboratory, Oxford University, Parks Road, Oxford OX1 3PU, UK}

\begin{document}

\def\papertitle{{Absolutely Stable Spatiotemporal Order in Noisy Quantum Systems}}
\def\authornames{{Max McGinley, Sthitadhi Roy, and S.~A.~Parameswaran}}

\title{\papertitle}
\author{Max McGinley}
\affiliation{\ox}
\author{Sthitadhi Roy}
\affiliation{\ox}
\affiliation{Physical and Theoretical Chemistry, Oxford University, South Parks Road, Oxford OX1 3QZ, UK}
\affiliation{International Centre for Theoretical Sciences, Tata Institute of Fundamental Research, Bengaluru 560089, India}
\author{S.~A.~Parameswaran}
\affiliation{\ox}
\date{\today}

\begin{abstract}
We introduce a model of non-unitary quantum dynamics that exhibits infinitely long-lived discrete spatiotemporal order robust against any unitary or dissipative perturbation. Ergodicity is evaded by combining a sequence of projective measurements with a local feedback rule that is inspired by Toom's `North-East-Center' classical cellular automaton. The measurements in question only partially collapse the wavefunction of the system, allowing some quantum coherence to persist. 
{We demonstrate our claims using numerical simulations of a Clifford circuit in two spatial dimensions which allows access to large system sizes, and also present results for more generic dynamics on modest system sizes.}
We also
devise explicit experimental protocols realising this dynamics using one- and two-qubit gates that are available on 
present-day quantum computing platforms.
\end{abstract}

\maketitle

\textit{Introduction.---}Quantum systems
driven out of equilibrium can exhibit phenomena that have no equilibrium analogues. 
This has driven a recent thrust aimed at identifying intrinsically nonequilibrium phases of matter in isolated periodically-driven quantum many-body systems, which include discrete time crystals (DTCs) \cite{Khemani2016,Keyserlingk2016a, Else2016,moessner2017equilibration,yao2017discrete} and Floquet topological phases \cite{Rudner2013,Potter2016,Else2016a,Keyserlingk2016,Roy2017}. Each of these phases is characterized by a particular pattern of spatiotemporal order that is infinitely long-lived in the thermodynamic limit, thus necessitating a mechanism for avoiding ergodicity. In locally interacting systems with unitary dynamics, this can be achieved by using spatial disorder to encourage many-body localization (MBL) \cite{Nandkishore2015}.

The high degree of control offered by quantum simulators makes them a promising platform for experimentally realising such phases, and progress is already being made in this direction \cite{Ippoliti2021a, Mi2021}. However, with current technology, these platforms inevitably
suffer from noise, which destabilizes MBL \cite{Nandkishore2014,Johri2015}. Therefore, in sufficiently large systems the lifetime of temporal ordering will be noise-limited. This presents a question: given a locally interacting noisy quantum system, what kinds of temporal order can be realised  that are 1) infinitely long-lived in the thermodynamic limit, and 2) robust against weak unitary and dissipative perturbations?

Here, we devise a scheme to realise a DTC using strictly local interactions that satisfies the above two criteria. Instead of MBL, projective measurements
are combined with conditional feedback 
to evade ergodicity. The measurements we propose do not fully collapse the wavefunction of the system, thus
admitting genuinely quantum dynamics. We show that the resulting spatiotemporal DTC order is absolutely stable against all perturbations \cite{Keyserlingk2016a}, including those that break symmetries, and that their correlation time scales exponentially with system size. We
devise a protocol to implement this scheme on present-day superconducting quantum processors. 

Many-body dynamics
in the presence of  projective measurements have been extensively explored,
particularly with regard to entanglement transitions \cite{Li2018,Li2019,Bao2020,Gullans2020,Fan2020,Ippoliti2021}. This generates ensembles of measurement outcomes $w$ and corresponding states $\rho_w$ that  can exhibit various orders \cite{Lavasani2021,Sang2021,Bao2021}; however to probe this ensemble experimentally is exponentially hard. Rather than considering the measurement-conditioned states, we use the outcomes of measurements to influence the dynamics itself, in a way that counteracts the deleterious effects of noise.

The DTC we consider here should be distinguished from proposals that use long-range interactions \cite{Russomanno2017,Choi2017,Gong2018,Gambetta2019}, fine-tuned dynamical symmetries \cite{Buca2019,Chinzei2020}, or macroscopically occupied bosonic modes \cite{Smits2018,Pizzi2019} to suppress fluctuations, as well as those in zero-dimensional systems \cite{Sacha2015,Gong2018}. Indeed, the spatial dimension of the system, which determines connectivity, plays a key role here. We argue that a measurement-feedback-stabilized DTC is only implementable using local interactions in spatial dimension $d \geq 2$  (without requiring a prohibitively large local state space). This can be compared to an analogous observation for systems subject to time-dependent Lindblad dynamics \cite{Lazarides2020}, as well as classical systems \cite{Yao2020, Zhuang2021}; we discuss connections between these various protocols.

\textit{Measurement-feedback stabilized time crystal.---}
We consider a system of $N$ qubits on a regular $d$-dimensional lattice whose dynamics respects discrete time-translation symmetry, i.e.~the evolution repeats itself after a given time period. The evolution of the density matrix over one such period (which we set to unity) is captured by a quantum channel $\mathcal{N}$ (a completely positive trace-preserving map) such that $\rho(t+1) = \mathcal{N}[\rho(t)]$ (we fix the period to unity). $\mathcal{N}$ plays an analogous role to the Floquet unitary in isolated systems. We split the evolution into two steps: $\mathcal{N} = \mathcal{N}_2 \circ \mathcal{N}_1$. In the first, the qubits are subjected to single-qubit rotations
\begin{align}
\mathcal{N}_1[\rho] &= U_1 \rho U_1^\dagger & U_1 = \prod_j e^{-\iu \theta_j Z_j/2} \label{eq:ChannelPulse}
\end{align}
where $Z_j$ is the third Pauli matrix acting on qubit $j$. The variation of $\theta_j$ between qubits allows us to describe unintended deviations from some desired pulse angle $\bar{\theta}$, which we presume to be small $|\theta_j - \bar{\theta}| \ll 1$. Later, we will also include incoherent errors during this step.

If $\theta_j = \pi$ for all $j$ then any qubit in the state  $\ket{\pm} \coloneqq (\ket{0} \pm \ket{1})/\sqrt{2}$ will evolve to $\ket{\mp}$ under \eqref{eq:ChannelPulse}, giving rise to oscillations of the magnetization $\langle M\rangle \coloneqq N^{-1} \langle \sum_j X_j \rangle$ with period 2. However, deviations of $\theta_j$ from $\pi$ will clearly destroy this subharmonic response in the long time limit. One way to stabilize these otherwise fine-tuned oscillations is to include a second unitary step $\mathcal{N}_2[\rho] = U_2 \rho U_2^\dagger$ where $U_2$ features strong spatial disorder, driving the system into a Floquet-MBL phase \cite{Khemani2016, Else2016}. Such an MBL-DTC is robust against weak perturbations, provided that the dynamics remains unitary and exactly time-periodic \cite{Keyserlingk2016a}. In our case, the stabilization step $\mathcal{N}_2$ is intrinsically nonunitary, and this leads to a DTC of a fundamentally different character. To illustrate how this can be achieved, we first discuss a $d = 1$ setup that fails to fully stabilize a DTC, and then provide a $d = 2$  protocol that succeeds and, \textit{inter alia}, explain the dimensional distinction.

In $d = 1$, a seemingly useful strategy is to perform projective measurements of domain wall operators $W_j \coloneqq X_j X_{j+1}$. Domain wall measurements provide us with information about the defects incurred by imperfections in $\mathcal{N}_1$, and based on this information we aim to apply operations that systematically remove errors. As mentioned above,
we will insist that this feedback be \textit{local}, i.e.~we decide what operation to apply to qubit $j$ based only on
measurements within some fixed finite distance $r$ from $j$. Unfortunately, all simple \footnote{There may exist a feedback strategy in $d = 1$ that mimics the dynamics of Ga\c{c}s' classical automaton \cite{Gacs1986, Gray2001}; however the only known protocols of this kind require a local state space of dimension order $2^{400}$, which we rule out as infeasible.} local strategies of this kind fail to simultaneously stabilize two independent steady states in $d = 1$, leading to loss of temporal correlations. The reason is that domain walls can only be eliminated in pairs, making isolated domain walls uncorrectable. Without fine-tuning, there is a nonzero probability that nearby domain walls will separate by a distance $>r$ before they are detected, whereupon they cannot be removed by a local rule. A similar observation has been made for Lindblad dynamics \cite{Lazarides2020}.


\begin{figure}
	\includegraphics[width=246pt]{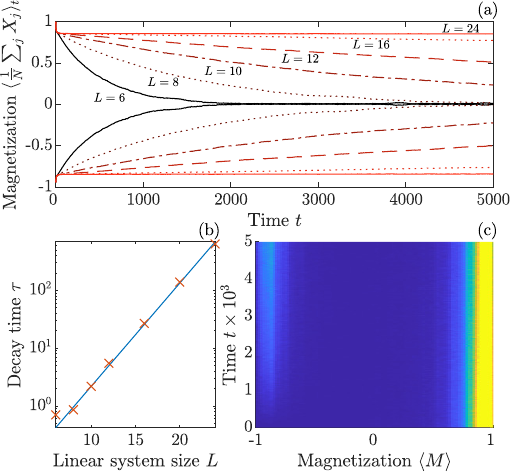}
	\caption{Oscillations of the  magnetization, averaged over 
	$10^4$ trajectories, under noisy Clifford dynamics initialized with  $\rho(0) = \bigotimes_j \ket{+}\bra{+}_j$. (a) Dynamics of magnetization for varying $L$; even and odd times plotted separately. (b)
	Estimated decay times $\tau$, extracted by fitting even-time data from (a) to an exponential,  plotted
	against $L$ (red crosses). We find  $\tau\propto e^{L/\xi}$ (blue line). (c) Histogram of the even-time sample magnetization
	for $L = 12$.
	We fixed 
	$(p_{\rm flip}, p_{\rm NEC}, p_{\rm unit}, p_{\rm reset}, p_{\rm ME}) = (0.95, 0.8, 0.02, 0.02, 0.01)$.}
	\label{fig:Oscillations}
\end{figure}

Instead, we must turn to $d = 2$, where domain walls are line-like, rather than point-like. Here we \textit{can} define a local feedback rule that successfully removes errors by encouraging closed loops of domain walls to shrink. The rule is closely related to Toom's `North-East-Centre rule' (NEC) classical cellular automaton, which has
provably robust bistability \cite{Toom1980}. In the NEC rule, defined on the square lattice, each classical spin is flipped whenever its North and East neighbours are both opposite to itself. Under this update rule, any small domain wall loop will shrink in a Southwesterly direction, favouring uniformly ordered states (all 0s or all 1s). Our protocol constitutes a quantum version of the NEC rule: For a given qubit we measure domain wall operators $W_{j, \hat{a}} = X_j X_{j+\hat{a}}$, where $\hat{a}$ is either $\hat{n} = (1,0)$ or $\hat{e} = (0,1)$; if both have outcome $-1$, we apply a $\pi$ pulse $e^{-\iu \pi Z_j/2}$. Formally, this operation, which acts on the three qubits $j, j+\hat{n}, j + \hat{e}$ can be described by a quantum channel \begin{align}
	\mathcal{N}_{{\rm T}, j}[\rho] = \hspace{-10pt}\sum_{w_{\hat{n}}, w_{\hat{e}} = \pm 1}
	U_{w_{\hat{n}}, w_{\hat{e}}} 
	\Pi_{j, \hat{n}}^{w_{\hat{n}}}\,\Pi_{j, \hat{e}}^{w_{\hat{e}}}\, \rho\, \Pi_{j, \hat{e}}^{w_{\hat{e}}}\, \Pi_{j, \hat{n}}^{w_{\hat{n}}}
	U_{w_{\hat{n}}, w_{\hat{e}}}^\dagger 
	\label{eq:ToomQubitChannel}
\end{align}
where $\Pi_{j, \hat{a}}^{w_{\hat{a}}} = (1 + w_{\hat{a}}W_{j, \hat{a}})/2$ projects onto the eigenspace of $W_{j, \hat{a}}$ with eigenvalue $w_{\hat{a}}$, and $U_{w_{\hat{n}}, w_{\hat{e}}}$ is the conditional unitary, which equals $e^{-\iu \pi Z_j/2}$ if $w_{\hat{n}} = w_{\hat{e}} = -1$, and $\mathbbm{I}$ otherwise. This channel outputs a density matrix that is a classical mixture of all possible outcomes. In principle, we could  examine the full ensemble of states conditioned on measurement outcomes by `unravelling' the channel \cite{Gullans2020} and examining the ensemble $\{(p_{\vec{w}}, \rho_{\vec{w}})\}$, where $\vec{w}$ is a measurement history, $p_{\vec{w}}$ is its probability, and $\rho_{\vec{w}}$ is the state conditioned on $\vec{w}$. However, sampling the full space of measurement outcomes is exponentially hard, so we instead focus on the mixture \eqref{eq:ToomQubitChannel}.

In the full stabilization step $\mathcal{N}_2$, we choose to apply the measurement-feedback sequence $\mathcal{N}_{{\rm T}, j}$ to each qubit in a particular sublattice $j \in A$ independently with probability $p_{\rm NEC}$; we then do the same for the opposite sublattice $B$. (Note that $\{\mathcal{N}_{{\rm T}, j}\}$ do not necessarily commute on different sublattices.). Formally, we have $\mathcal{N}_2 = \prod_{j \in B} [(1-p_{\rm NEC}) + p_{\rm NEC}\mathcal{N}_{{\rm T}, j}] \prod_{j \in A} [(1-p_{\rm NEC}) + p_{\rm NEC}\mathcal{N}_{{\rm T}, j}] $. The maximum error correction rate occurs at $p_{\rm NEC} = 1$.


\textit{Numerical simulations.---} 
We now test these ideas using numerical simulations in two ways. First, we employ Clifford circuitry, which involves a restricted range of operations that can be simulated efficiently for large system sizes \cite{STIM}. The allowed unitary and nonunitary operations are sufficiently diverse, allowing us to verify the robust nature of the spatiotemporal order. Second, for modest $L$ we simulate circuits with arbitrary gates; these results are described in the supplement~\cite{SM}.

The full sequence of operations in the Clifford circuits is as follows. In place of partial rotations $\theta_j \neq \pi$, we apply $\pi$ pulses to each spin with some probability $p_{\rm flip}$. Then, for each qubit $j$, a two-qubit gate $e^{-\iu \pi Z_j Z_{j'}/4}$ is applied with probability $p_{\rm unit}$ for some randomly chosen neighbour $j'$. Qubits are then reset into the $\ket{+}_j$ at random with probability $p_{\rm reset}$ (note that this explicitly breaks the Ising $+X_j \leftrightarrow -X_j$ symmetry). Finally, the correction step $\mathcal{N}_2$ is made. We model measurement errors by inverting measurement outcomes with probability $p_{\rm ME}$ before deciding whether or not to apply the correcting $\pi$-pulse. In all data, averages over the  measurement outcomes and  random decisions  are performed simultaneously.

To probe DTC order, we initialize the system in an $X_j=1$ eigenstate
for all $j$, and calculate the expectation value of the magnetization $\braket{M}_t= \braket{\frac{1}{N} \sum_j X_j}$ after $t$ timesteps. For nonzero but sufficiently small error probabilities ($p_{\rm unit},\,p_{\rm reset},\,p_{\rm ME})$, we see the hallmark period-doubled oscillations [Fig.~\ref{fig:Oscillations}(a)], with an amplitude that quickly reaches a quasi-stationary value after some $O(1)$ time.  In finite-size systems the oscillations eventually decay at late times, but the timescale of this decay $\tau$ grows exponentially with the linear system size $L$ [Fig.~\ref{fig:Oscillations}(b)]. The distribution of magnetization $M$ [Fig.~\ref{fig:Oscillations}(c)] remains bimodal, implying that the decay is due to rare events where the sign of the magnetization flips across the entire system. Such processes require domain walls to traverse the system before being corrected, which requires $O(L)$ spin-flips in an $O(1)$ time, explaining the dependence of $\tau$ on $L$. Thus, despite the presence of noise, these oscillations are infinitely long-lived in the thermodynamic limit. We have confirmed that the same oscillations are seen for generic, partially magnetized initial states \cite{SM}, and when the two-qubit unitaries are replaced by random Clifford gates, and decoherence in the $Z$ basis is included.

A more rigorous way to identify a time crystal is to look for spatiotemporal order \cite{Khemani2017}, conveniently probed by the correlator $C_t(j,j') \coloneqq \braket{X_j(t) X_{j'}(0)}$, where the operator $X_j(t) = (\mathcal{N}^\dagger)^t[X_j]$ evolves in the Heisenberg picture, and the expectation value is taken with respect to a steady state of the dynamics. We have confirmed that $C_t(j,j')$ approaches a nonzero value  whose value oscillates with period 2 as $|j-j'| \rightarrow \infty$ \cite{SM}, indicating that the characteristic DTC 
order is robust in the thermodynamic limit.

\begin{figure}
	\includegraphics[width=246pt]{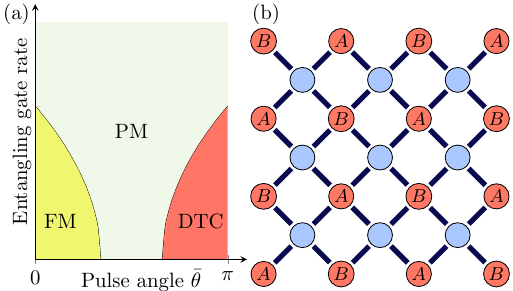}
	\caption{(a) Qualitative phase diagram as a function of the mean pulse angle $\bar{\theta}$ [Eq.~\eqref{eq:ChannelPulse}] and 
	the strength of entangling unitary evolution. We distinguish the paramagnetic phase (PM), where the system equilibrates and the steady state is unique, from the Ferromagnetic (FM) and discrete time-crystalline (DTC) phases, which in the thermodynamic limit are nonergodic, with persistent oscillations in the latter case. The DTC exhibits period-doubled oscillations in magnetization and autocorrelators, while these quantities reach a time-independent value in the FM. (b) Geometry of a superconducting processor that realises the measurement-feedback stabilized DTC. System qubits (red circles) that exhibit  DTC order are capacitively coupled (solid lines) to ancillas (blue circles), facilitating the necessary measurements.}
	\label{fig:PhaseDiagExp}
\end{figure}

The same quantities can be used to identify other phases in the parameter space. If the pulse angles $\theta_j$ are close to 0 rather than $\pi$ (or the flip rate $p_{\rm flip}$ is small), then the magnetization will not oscillate, instead reaching a static value. When noise is sufficiently weak, this saturation value depends on the initial state, which implies that the time evolution channel $\mathcal{N}$ has multiple steady states in the thermodynamic limit, spanned by two density matrices $\rho_{{\rm ss}, +}, \rho_{{\rm ss}, -}$ that correspond to opposite signs of magnetization. We refer to this phase as a ferromagnet (although unlike conventional ferromagnets, this behaviour is robust against perturbations that break the Ising symmetry, e.g.~resets in the Clifford circuits described above). If noise is increased, or the correction rate $p_{\rm NEC}$ is reduced, then eventually this bistability is lost, leading to a paramagnetic phase where the magnetization reaches a static, initial-state-independent value. A qualitative phase diagram is shown in Fig.~\ref{fig:PhaseDiagExp}(a).

Transitions between adjacent phases (PM to FM or DTC) can be driven by a number of different parameters. An example of particular interest is the entangling unitary gates, occuring with probability $p_{\rm unit}$. The scrambling nature of these processes encourages internal thermalization, where local subsystems equilibrate by becoming entangled with the rest of the system; thus we expect that chaotic unitary evolution competes with the nonergodic FM and DTC phases. We find that the system remains nonergodic up to a finite value of $p_{\rm unit}$, where a transition to the PM phase occurs. We find critical behaviour consistent with the Ising universality class~\cite{SM}.
{Note that the DTC-PM transition in a noisy, driven, classical spin system is in the same universality class~\cite{Gambetta2019a}, despite the DTC being a genuine nonequilibrium state}.


\textit{Experimental implementation.---} Finally, we demonstrate that our DTC can be experimentally realised using resources that are currently available in most superconducting qubit quantum computers with two-dimensional architectures. Our proposal requires single-qubit rotations and measurements combined with one entangling gate, which we take to be CNOT. (Gates equivalent to CNOT up to single-qubit unitaries, including CZ \cite{Foxen2020,Arute2019} and cross-resonance \cite{Chow2011,Patterson2019} also suffice.)

%

Qubits are arranged on a square lattice with nearest-neighbour connectivity [Fig.~\ref{fig:PhaseDiagExp}(b)].
The `system'  qubits on one sublattice (dark red) 
will exhibit DTC order, and are coupled to ancillas
on the opposite sublattice (light blue). The system sites form a larger square lattice; 
evidently, any pair of neighbouring system qubits 
are coupled to least one shared ancilla
. To measure $W_{j, \hat{n}}$, we prepare ancilla $b$ that is connected to both $j$ and $j+\hat{n}$ in the state $\ket{+}_b$, and apply  $\text{CNOT}_{b \rightarrow j}$, 
followed by $\text{CNOT}_{b \rightarrow j+\hat{n}}$ ($\text{CNOT}_{c\rightarrow t}$ is a CNOT with $c, t$ as control and target, respectively). Finally the ancilla is measured in the $X_b$ basis. 
We may verify~\cite{SM} that an outcome $X_b = \pm 1$ projects the state onto the subspace $W_{j, \hat{n}} = \mp 1$, as desired. The same procedure can be used to measure $W_{j,\hat{e}}$, and the feedback gate can then be applied if both outcomes are $-1$, thus simulating the channel \eqref{eq:ToomQubitChannel}. 

Our simulations so far have featured periodic boundary conditions, however this is not always possible to implement in experiments. The NEC rule will sometimes fail to correct errors if na{\"i}vely generalized to open boundary conditions \cite{Vasmer2021}, due to corners without North or East neighbours. The simplest remedy is to switch to a majority vote rule, where all four domain wall operators adjacent to a site $j$ are measured and the $\pi$ pulse is applied if at least two domain walls are present. The corresponding classical cellular automaton exhibits robust bistability with open boundaries; thus we expect that such a quantum feedback rule will stabilize a DTC. Unlike the NEC rule, the majority vote respects detailed balance, suggesting that the error elimination mechanism will be slower \cite{Bennett1985}, and that true bistability may be compromised in the ferromagnetic phase if the $X_j \rightarrow -X_j$ $\mathbb{Z}_2$ symmetry is broken \cite{Bennett1990}. (In the DTC phase, any up-down bias incurred in one timestep is cancelled out in the next.) Alternatively, we may retain the advantages of the NEC rule with open boundaries by employing an unusual annular geometry~\cite{SM}.

Projective measurements of superconducting qubits are typically slow compared to gate times. In IBM's current cloud-based quantum devices, the readout time is $\sim \SI{5}{\micro\second}$, while the $T_1$, $T_2$ times are $\sim \SI{100}{\micro\second}$ \cite{IBMQ}. Thus, the effective error rate per layer will of order $5\%$. Using Clifford circuits, we have confirmed that a DTC can be stabilized even with depolarizing errors as high as this, assuming measurement errors of $1\%$ and the same $p_{\rm NEC} = 0.8$ as in Fig.~\ref{fig:Oscillations}. As an alternative, in the supplement \cite{SM} we suggest a method to implement the same channel \eqref{eq:ToomQubitChannel} using purely quantum gates and qubit resets, the latter of which can be done much faster ($\SI{250}{\nano \second}$ in Ref.~\cite{Mcewen2021}).

\textit{Discussion.---}
In closing, it is helpful to compare our model with analogous systems that use engineered Lindblad dynamics as an entropy drain for stabilizing DTCs \cite{Lazarides2020}. Although Lindbladians are defined on continuous rather than discrete time, it is possible to qualitatively compare the two by considering the Lindblad dynamics using quantum jump trajectories \cite{Dalibard1992}, built up of Poisson-distributed discrete `jump events'. The jump operators in Ref.~\cite{Lazarides2020} are chosen according to a majority vote with all four neighbours. One example is $L_{j, \vec{b}} = \ket{b_j b_n b_e b_s b_w}\bra{\bar{b}_j b_n b_e b_s b_w}$, where $b \in \{+, -\},\, \bar{b} \coloneqq - b$, and $n, e, s, w$ are the North, East, South, and West neighbours of $j$, respectively.
After a jump event $\ket{\psi} \rightarrow L_{j, \vec{b}}\ket{\psi}$, the state of the five qubits involved is projected into a product of $X_j$ eigenstates. This differs in a crucial way from the measurement-feedback loop in our model: Domain wall operators are measured rather than individual qubit $X_j$ operators, which means that some quantum coherence can be preserved. For instance, if the three qubits $c, e, n$ involved in $\mathcal{N}_{{\rm T}, j}$ [Eq.~\eqref{eq:ToomQubitChannel}] begin in the superposition state $\alpha \ket{-_c +_e +_n} + \beta \ket{+_c -_e -_n}$, then the output state will be $\alpha \ket{+_c +_e +_n} + \beta \ket{-_c -_e -_n}$, which is also coherent. As a result, the steady states of the channel $\mathcal{N}_2$ form a coherent subspace: Any state within the Bloch sphere spanned by the pure states $\ket{+^{\otimes N}},\, \ket{-^{\otimes N}}$ is unaffected by $\mathcal{N}_2$, which implies the existence of a decoherence-free subspace \cite{Lidar1998}. In contrast, if we measured $X_j$ operators, or used the Lindblad model of Ref.~\cite{Lazarides2020}, then only incoherent mixtures $(1-p)\ket{+^{\otimes N}}\bra{+^{\otimes n}} + p\ket{-^{\otimes N}}\bra{-^{\otimes n}}$ would be stabilized. (For a more detailed comparison of the two kinds of dynamics, it is useful to take a continuum time limit of our model
~\cite{SM}.) Other practical differences are that the discrete measurement-feedback process needn't be active during the pulse step $\mathcal{N}_1$; also, in most experimental platforms with single-site control, conditional feedback is easier to achieve  than the reservoir engineering required by the Lindblad approach.

In practice, the massively multipartite entanglement contained in macroscopic coherent superpositions (`cat states') will be susceptible to noisy perturbations. Thus, we expect that classical information (the sign of the initial magnetization) will preserved for arbitrarily long [`$T_1$'$\rightarrow \infty$] times, while quantum information will persist over a noise-limited [`$T_2$']  timescale. In the language of quantum error correction, $\mathcal{N}_2$ constitutes a strictly local implementation of a repetition code \cite{Nielsen2010}, and so either $X$-type or $Z$-type errors can be corrected, not both. Other classical cellular automata have also been used as a basis for local implementations of different quantum codes \cite{Harrington2004}.

Very recent work  has leveraged the NEC rule  to realise absolutely stable time-crystalline order in classical Hamiltonian systems with Langevin noise \cite{Zhuang2021}. Our results show that reliable classical automata such as the NEC rule are also stable against quantum fluctuations, generated by e.g.~entangling unitary gates, up to some finite rate [see Fig.~\ref{fig:PhaseDiagExp}(a)]. The bistability of the classical automaton in question translates to the existence of robust multiple steady states of the quantum channel $\mathcal{N}_2$. Each steady state forms a basin of attraction for the dynamics \cite{Gambetta2019}, and the pulses in $\mathcal{N}_1$ map states in the basin of one attractor to that of the other.

Our study highlights the potential of measurement-feedback loops for synthesizing interesting kinds of nonunitary dynamics, which can be probed experimentally without issues of scalability. This strategy can be seen as a useful alternative to reservoir engineering \cite{Poyatos1996} that is particularly appropriate for systems with single-qubit control. We look forward to investigating the interplay between this paradigm of dynamics and other kinds of quantum nonequilibrium phases of matter.

\vspace{0.07in}
\begin{acknowledgments}
\textit{Acknowledgements.---} We thank A.~Daley for discussions, and M.~Ippoliti for helpful comments on the manuscript. We acknowledge support from  UK Engineering and Physical Sciences Research Council  grant EP/S020527/1. Statement of compliance with EPSRC policy framework on research data: This publication is theoretical work that does not require supporting research data. SR also acknowledges support from an ICTS-Simons ar
\end{acknowledgments}

\bibliography{toom_tc.bib}


\newpage
\afterpage{\blankpage}

\newpage
\begin{onecolumngrid}


\begin{center}
	{\fontsize{11}{11}\selectfont
		\textbf{Supplemental Material:  ``\papertitle''\\[5mm]}}
{\normalsize \authornames\\[1mm]}
	
\end{center}
\normalsize\
\end{onecolumngrid}

\begin{twocolumngrid}

\setcounter{equation}{0}
\setcounter{figure}{0}
\setcounter{table}{0}
\setcounter{page}{1}

\renewcommand{\theequation}{S\arabic{equation}}
\renewcommand{\thefigure}{S\arabic{figure}}

\subsection*{Continuum time limit}

In this section, we discuss in detail the continuum-time limit described in the main text, and compare this to the time-dependent Lindblad dynamics studied in Ref.~\cite{Lazarides2020}.

The correction step $\mathcal{N}_2$ involves projective measurements, which are intrinsically discontinuous events. Nevertheless, a continuum time limit can be taken by applying the three-qubit correction processes $\mathcal{N}_{{\rm T}, j}$ [Eq.~\eqref{eq:ToomQubitChannel}] according to independent random Poisson processes for each $j$, with some correction rate $\Gamma$. Specifically, in an infinitesimal time window $\dif t$, the channel $\mathcal{N}_{{\rm T}, j}$ is applied with probability $\Gamma \dif t$ for each $j$, and we do nothing otherwise (with probability $1-N\Gamma \dif t$). Since the Poisson processes are uncorrelated in time, the resulting dynamics can be described by a Markovian Lindblad master equation $\dif \rho/\dif t = \mathcal{L}[\rho]$ with time-independent superoperator $\mathcal{L}$, which without loss of generality can be cast in the standard diagonal form \cite{Breuer2002}
\begin{align}
    \mathcal{L}[\rho] = -\iu [H, \rho] +  \sum_\mu L_\mu \rho L_\mu^\dagger - \frac{1}{2} \{L^\dagger_\mu L_\mu, \rho \},
\end{align}
The operators $\{L_{\mu}\}$ are referred to as jump operators, and $H$ is Hermitian, playing the role of an effective Hamiltonian.

Evidently, the Lindbladian $\mathcal{L}$ pertaining to the stochastic dynamics described above is given by
\begin{align}
    \mathcal{L} = \lim_{\substack{p, \dif t \rightarrow 0 \\ p = \Gamma \dif t}}  \frac{(1-Np){\rm id} + p \sum_j \mathcal{N}_{{\rm T}, j}  }{\dif t}
\end{align}
where $\rm{id}$ is the identity superoperator. Using this expression, one can readily determine that $H = 0$, and that there are four jump operators $\{L_{j,m} : m = 1, 2, 3, 4\}$ associated with each site $j$
\begin{align}
    L_{j, m} &= \sum_{b = \pm} L_{j, m, b} \label{eq:JumpOperatorToom} \\
    L_{j, 1, b} &= \sqrt{\Gamma} \ket{b_j b_{j+\hat{e}} b_{j+\hat{n}}}\bra{b_j b_{j+\hat{e}} b_{j+\hat{n}}} \tag{\theequation a}\\
    L_{j, 2, b} &= \sqrt{\Gamma} \ket{b_j \bar{b}_{j+\hat{e}} b_{j+\hat{n}}}\bra{b_j \bar{b}_{j+\hat{e}} b_{j+\hat{n}}} \tag{\theequation b}\\
    L_{j, 3, b} &= \sqrt{\Gamma} \ket{b_j b_{j+\hat{e}} \bar{b}_{j+\hat{n}}}\bra{b_j b_{j+\hat{e}} \bar{b}_{j+\hat{n}}} \tag{\theequation c}\\
    L_{j, 4, b} &= \sqrt{\Gamma} \ket{b_j b_{j+\hat{e}} b_{j+\hat{n}}}\bra{b_j \bar{b}_{j+\hat{e}} \bar{b}_{j+\hat{n}}} \tag{\theequation d}
\end{align}
where $\ket{b_j b_{j+\hat{e}} \cdots}$ is shorthand for $(\ket{0}_j + b_j \ket{1}_j)/\sqrt{2} \otimes  (\ket{0}_{j+\hat{e}} + b_j \ket{1}_{j+\hat{e}})/\sqrt{2} \otimes \cdots$; we write $\bar{b}$ to denote $-b$; and as usual $j + \hat{e}$, $j + \hat{n}$ are the Eastern and Northern neighbours of qubit $j$. Each of the four jump operators comes from a particular term in the sum in Eq.~\eqref{eq:ToomQubitChannel}, which correspond to the different outcomes that could occur when measuring both domain wall operators. The Lindbladian can be supplemented with a (possibly time-dependent) Hamiltonian to describe a driving field that effects the oscillation of magnetization (in place of the discrete pulse channel $\mathcal{N}_1$), as well as additional jump operators that model unintended dissipative effects.

Now that the dynamics is cast in a Lindblad form, it is possible to compare directly with that of Ref.~\cite{Lazarides2020}. There, the jump operators act on a given spin $j$ and its four North, East, South, and West neighbours. Each jump operator has the form
\begin{align}
    L_{j, \vec{b}} = \sqrt{\Gamma} \ket{f(\vec{b})_j b_{j+\hat{e}}^{(e)} b_{j+\hat{n}}^{(n)} b_{j+\hat{w}}^{(w)} b_{j+\hat{s}}^{(s)} } \bra{b_j^{(c)} b_{j+\hat{e}}^{(e)} b_{j+\hat{n}}^{(n)} b_{j+\hat{w}}^{(w)} b_{j+\hat{s}}^{(s)}}
    \label{eq:JumpOperatorMajvote}
\end{align}
where $\vec{b}$ is a five-dimensional vector whose components $\{ b^{(c)}, b^{(e)}, b^{(n)}, b^{(w)}, b^{(s)}\}$ are 0 or 1. The function $f : \{0, 1\}^{\times 5} \rightarrow \{0, 1\}$ implements a majority vote of the 5 constituent spins.

The dynamics described above differs from ours in two important ways. Firstly, the jump operator \eqref{eq:JumpOperatorMajvote} updates the central spin depending on the majority of itself and its four neighbours, whereas in our case [Eq.~\eqref{eq:JumpOperatorToom}] only the Northern and Eastern neighbours affect the flip process. In essence, the isotropic majority vote rule induces local transitions according to whether or not they reduce the energy of the system with respect to the Ising Hamiltonian $H_{\rm Ising} = \sum_{\langle j, j'\rangle} X_j X_{j'}$; thus if no other terms are added the dynamics will obey detailed balance with respect to $H_{\rm Ising}$. In contrast, the jump operators \eqref{eq:JumpOperatorToom} found in our model do not satisfy detailed balance, since the interactions between neighbouring spins are strictly one-way ($j$ will flip depending on the state of $j+\hat{n}$, but the probability of $j+\hat{n}$ flipping is independent of the state of $j$). The intrinsically non-equilibrium nature of the NEC-based dynamics allows for truly robust bistability, whereas the equilibrium model can only exhibit bistability in regions of parameter space of measure zero -- specifically on subspaces where Ising symmetry is respected. (For a detailed discussion of the differences between classical automata based on reversible vs.~irreversible update rules, see e.g.~Refs.~\cite{Bennett1985, Grinstein2004}.)

Although the stabilization terms in Ref.~\cite{Lazarides2020} obey detailed balance, it is still possible to reach a bistable region if additional driving is added, e.g.~using a time-dependent Hamiltonian, which takes the system away from equilibrium. Indeed, one can argue that the majority vote jump operators \textit{can} stabilize a period-2 time-crystalline phase against symmetry-breaking perturbations: The effect of any bias favouring $+X$ magnetization over $-X$ (e.g.~a jump operator $L_{j, \rm bias} = \ket{+}\bra{-}_j$) acting over one time period will be cancelled out by itself during the next time period, since the sign of the magnetization is reversed each step \cite{Bennett1990}. Thus, we expect that the phase diagram of the model of Ref.~\cite{Lazarides2020} will feature a robust DTC phase, but the ferromagnetic phase that we found in our model will not be present, since any bias will not be cancelled out and will amplify over time. Moreover, the same strategy will not be able to stabilize a time crystal with period $n > 2$, whereas our NEC-model could in principle be generalized to realise higher-period time crystals.

The second key difference between the two models is that the jump operators \eqref{eq:JumpOperatorMajvote} are fully incoherent: Any jump event fully collapses the wavefunction of the five constituent spins to a product of $\pm X_j$ eigenstates. In contrast, as discussed in the main text, coherences in the $X_j$ basis are partially preserved under our update rule. This can be seen immediately from the form of the jump operators \eqref{eq:JumpOperatorToom}, each of which is a coherent sum of two operators that have identical domain wall configurations, but opposite magnetizations of each spin. An incoherent version of our rule could be devised where one would include each $L_{j, m, b}$ as a separate jump operator. Such a Lindbladian would differ from our model in that cross-terms such as $L_{j,m,+} \rho L_{j,m,-}^\dagger$ would not be present. Because of these cross-terms, any state within the Bloch sphere (coherent superpositions or classical mixtures) spanned by $\ket{+^{\otimes N}}$, $\ket{-^{\otimes N}}$ will be steady states of the coherent Lindbladian, whereas only classical mixtures are stabilized by the incoherent Lindbladian.

\subsection*{NEC rule with open boundary conditions}

In this section, we explain how to implement the NEC rule in an in-plane superconducting quantum processor without relying on periodic boundary conditions, which are not generally possible in experimental platforms. In brief, our strategy is to consider a NEC automaton on a torus embedded in 3-dimensional (3D) space (equivalent to a 2D system with periodic boundary conditions). We then `flatten' the system in one of the three-dimensional directions to obtain a quasi-2D system with top and bottom layers. These two layers are then merged into one, which results in a truly 2D system with an annular shape. This system inherits the robust non-ergodic properties of the NEC rule without requiring any cross-wiring, thus allowing for on-chip realizations.

\begin{figure*}
    \centering
    \includegraphics[width=510pt]{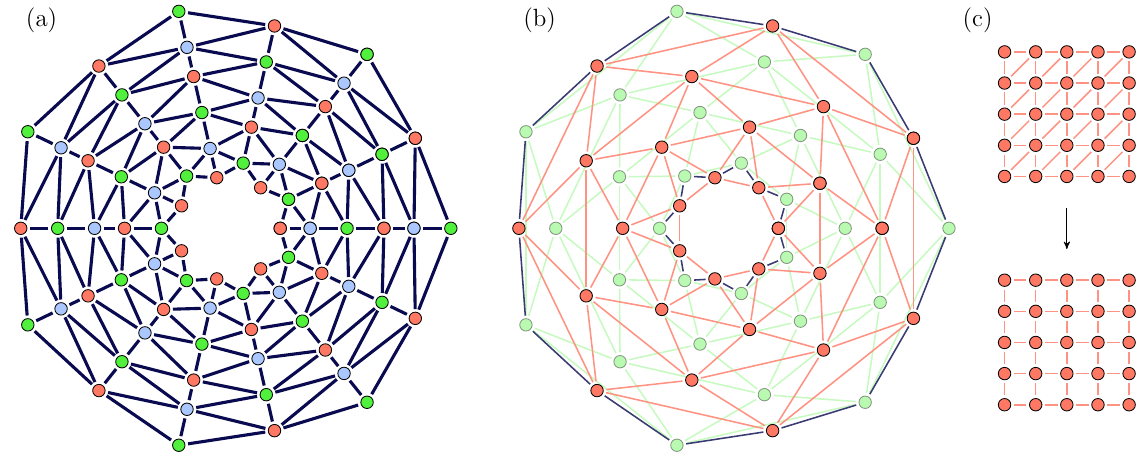}
    \caption{(a) Layout and connectivity of physical and ancilla qubits in a 2D annular geometry that can emulate a system with periodic boundaries. As in Fig.~\ref{fig:PhaseDiagExp}, blue circles are ancilla qubits, and dark blue lines represent capacitive couplings. The physical qubits are divided into two sublattices, coloured in red and green. Altogether, the qubits form a triangular lattice where one direction (radial) has open boundaries, and the other (polar) is effectively periodic (b) Effective connectivity of the system qubits. The qubits are drawn in the same positions, with ancillas removed. Any two physical qubits connected by a line have the property that there exists at least one ancilla qubit to which both are directly coupled, so that a domain wall operator between the two physical qubits can be measured (see main text). The red and green sublattices each form a triangular lattice. (The green sublattice is drawn faintly to aid the eye.) At the inner and outer boundaries of the annulus, the two sublattices are joined together (dark blue lines), such that the radial direction becomes effectively periodic. (c) A square lattice (bottom) can be obtained from a triangular lattice (top, rotated $45^\circ$ and stretched) by removing two out of six of the bonds coming out of each qubit.}
    \label{fig:annular}
\end{figure*}

In describing this construction in detail, we find it instructive to work backwards, starting from the strictly 2D open-boundary geometry and showing that its dynamics mimics that of a system with periodic boundaries. The layout and connectivity of the physical qubits (both system and ancilla) are illustrated in Fig.~\ref{fig:annular}(a). Each qubit is capacitively coupled to 6 neighbouring qubits in a triangular lattice structure. One-third of the qubits are ancillas, and the remaining two-thirds are split into two separate `system' sublattices, drawn in red and green, respectively.

Within each system sublattice, we must determine which domain wall operators $W_{j j'} = X_j X_{j'}$ can be projectively measured, using the technique described in the main text. This requires $j$ and $j'$ to be both connected to at least one ancilla qubit in common. The pattern of effective connectivities defined by this condition is illustrated in Fig.~\ref{fig:annular}(b) -- each sublattice now forms a smaller triangular lattice, with open boundary conditions in the radial direction and periodic boundary conditions in the azimuthal direction. At this point, we can `stitch together' the two sublattices at the inner and outer boundaries by adding connections between red and green qubits (these still satisfy the measurable domain wall criterion). We can then consider one sublattice (say red) to be the upper half of a torus, and the other to be the lower half. Specifically, in the standard two-angle parametrization of the torus $(\varphi, \theta)$, the red qubits reside on the region of poloidal angle $\theta \in [0, \pi)$, while the green qubits are mapped to $\theta \in [\pi, 2\pi)$.

This demonstrates that the 2D geometry in Fig.~\ref{fig:annular}(a) can mimic that of a triangular lattice on a torus. We now need to generalize the quantum NEC measurement-feedback rule, which is defined for the square lattice. This problem has been solved in Ref.~\cite{Kubica2019} for a broad class of (not necessarily regular) lattices with periodic boundary conditions. The `sweep rule' introduced there specifies an update rule for classical spins that, like the NEC rule, depends only on the presence or absence of domain walls on particular bonds, and gives rise to provably robust bistability. One must pick a sweep direction $\vec{h}$, which determines the way in which regions of errors shrink. In our case, it is convenient to pick this to be along the toroidal direction $\hat{e}_\varphi$, i.e.~the direction that corresponds to the polar coordinate of the annulus. With this choice, the sweep rule takes a particularly simple form, where one ignores two out of six of the connections emanating from each qubit, such that the remaining connections form a square lattice [see Fig.~\ref{fig:annular}(c)], and the NEC rule can be applied as usual. However, other choices will also be possible, provided that they are consistent with the requirements of the sweep rule. As before, the resulting classical automaton can be used to define a quantum measurement-feedback loop that inherits the same error-correcting properties.

Since the sweep direction $\vec{h}$ winds around the toroidal direction, it is not possible to `fill in' the hole at the centre of the annulus without encountering a point where the sweep direction has singular behaviour, which would cause the correcting mechanism to fail. If one wishes to have a simply connected 2D geometry, then one can repeat the trick of `squishing' the periodic direction into two copies of an open direction. The result would be a 2D system with four sublattices, each of which map onto one of the four quadrants of the torus ($(\varphi, \theta) \in [n \pi, (n+1)\pi) \times [m \pi, (m+1)\pi) $, for $n, m = 0, 1$). To do so will likely require higher connectivity of physical qubits, which may be challenging.

The annular embedding of the triangular lattice means that the distance between connected qubits grows as one moves away from the centre. Therefore, to make this design scalable with a fixed maximum qubit-qubit coupling distance, it may be necessary to insert extra qubits, forming disclinations of the triangular lattice. Since the sweep rule can be defined for lattices with such irregularities \cite{Kubica2019}, it will still be possible to devise a measurement-feedback protocol with the same robust properties.

One unusual artefact of the setup we have proposed is that any correlated two-qubit errors that act on pairs of system qubits on opposite sublattices will look highly non-local under the toric mapping. Fortunately, the measurement-feedback protocol remains robust against these kind of errors. We have simulated the 2D annular system, including unitary errors that act on any pairs of qubits that are physically connected, regardless of which sublattice they are on. We have confirmed that the characteristic exponential scaling of the correlation time is still seen.

\subsection*{Experimental domain wall measurement protocol}

Here we demonstrate explicitly that the protocol for measuring domain wall operators described in the main text does indeed project the state of the system onto one of the eigenspaces of $W_{j, j'}$. We assume that the two system qubits involved $j$, $j'$ begin in an arbitrary pure state $\ket{\psi}_{jj'}$, and the ancilla $b$ begins in the state $\ket{+}_b = (\ket{0}_b + \ket{1}_b)/\sqrt{2}$. The gate $U_{{\rm CR}, j, b, \pi/2} = e^{-\iu \pi X_j Z_b/4}$ is applied, followed by $U_{{\rm CR}, j', b, \pi/2}$. The resulting state is then
\begin{align}
    \ket{\Phi} &= e^{-\iu\pi(X_{j\vphantom{'}} + X_{j'})Z_b/4}\ket{\psi}_{jj'} \otimes \ket{+}_b \nonumber\\
    &= \Big[ \Pi_{X_{j\vphantom{'}} + X_{j'} = +2}  (-\iu Z_b) + \Pi_{X_{j\vphantom{'}} + X_{j'} = 0} \nonumber\\ &+ \hspace{6pt}\Pi_{X_{j\vphantom{'}} + X_{j'} = -2} (+\iu Z_b)\Big]  \ket{\psi}_{jj'} \otimes \ket{+}_b,
\end{align}
where in the last equality we have decomposed the unitary operator as the direct sum of its action on the different eigenspaces of the operator $(X_{j\vphantom{'}} + X_{j'})$, spanned by the projectors $\Pi_{X_{j\vphantom{'}} + X_{j'}} = \lambda$, for $\lambda = -2, 0, 2$. We have $Z_b \ket{+}_b = \ket{-}$, and also note that $\Pi_{X_{j\vphantom{'}} + X_{j'} = 0}$ is equal to the projector onto the eigenspace of the target operator $W_{j,j'} = X_j X_{j'}$ with eigenvalue $-1$, which we denote $\Pi_-$. The orthogonal projector onto the eigenspace $+1$ is equal to $\Pi_+ = \Pi_{X_{j\vphantom{'}} + X_{j'} = +2} + \Pi_{X_{j\vphantom{'}} + X_{j'} = -2}$, and so we identify
\begin{align}
    \ket{\Phi} &= e^{-\iu \pi X_j/2} \Pi_+ \ket{\psi}_{jj'} \otimes \ket{-}_b 
    + \Pi_-  \ket{\psi}_{jj'} \otimes \ket{+}_b,
    \label{eq:StatePostCR}
\end{align}
where we have used $e^{-\iu \pi X_j/2} \Pi_{X_{j\vphantom{'}} + X_{j'} = \pm 2} = \mp \iu \Pi_{X_{j\vphantom{'}} + X_{j'} = \pm 2} $. From this expression, we see that if the ancilla is measured in the $X_b$ basis, then an outcome of $+1$ will project the system qubits onto the space with a domain wall present, spanned by $\Pi_-$. The opposite outcome of $+1$ projects the system qubits onto the orthogonal space, with no domain wall present, with an additional single-qubit $X_j$-$\pi$-pulse applied. This extra gate can be explicitly corrected whenever this measurement occurs, although this is not actually necessary to observe time-crystalline order, since the $X_j$ gate does not disturb the key correlations in the system.

If the native gates of the quantum simulator in question are different from the cross-resonance gate, then it may still be possible to measure the domain wall operators using the same principle. Many commonly-used two qubit gates are equivalent to $U_{{\rm CR}, j, k, \theta}$ when conjugated with single-qubit unitaries. For instance, if one has access to a CPHASE gate $U_{{\rm CPHASE}, j, k, \theta} = e^{-\iu \theta (1-Z_j)(1-Z_k)/4}$ (or a CZ gate), then one can apply Hadamard gates to the system qubits $j$, $k$ before applying $U_{{\rm CPHASE}, j, b, \pi} U_{{\rm CPHASE}, j', b, \pi}$. Measuring the ancilla qubit in the $X_b$ basis then reveals the presence or absence of a domain wall, as before, and finally Hadamard gates can be applied again to counteract the initial rotations. 

Finally, we note that a slightly modified approach to implementing the channel \eqref{eq:ToomQubitChannel} can be constructed that removes the need for projective measurements, which are typically very slow compared to gate times, at the expense of requiring more two-qubit gates. The cross-resonance gates described above are applied to the $(j, j+\hat{e})$ bond using one ancilla $b_e$, and the same for the $(j, j+\hat{n})$ bond using a different ancilla $b_n$ [this is possible given the connectivity of the lattice in Fig.~\ref{fig:PhaseDiagExp}(b)]. Then, rather than measuring the two ancillas separately and acting on the central qubit $j$, one can instead implement a three-qubit quantum gate that has the effect of applying a $Z$ gate to the central qubit if both ancilla qubits are in the state $\ket{+}$. This gate is equivalent to a Toffoli gate up to single-qubit rotations, and so can be built up of several two-qubit unitaries \cite{Sleator1995}. Afterwards, the ancilla qubits can both be reset and prepared in the state $\ket{+}_b$ again, so that they are ready for the next timestep. Overall, this is potentially a much faster process, since the non-unitary processes involved (qubit resets) can be implemented more rapidly than measurements \cite{Mcewen2021}.

\subsection*{Initial States and Autocorrelator}

Here we provide results of additional numerical simulations that probe the initial-state dependence of the magnetization dynamics, the behaviour of the autocorrelator, and the critical dynamics of the measurement-feedback model.

\begin{figure}
    \centering
    \includegraphics{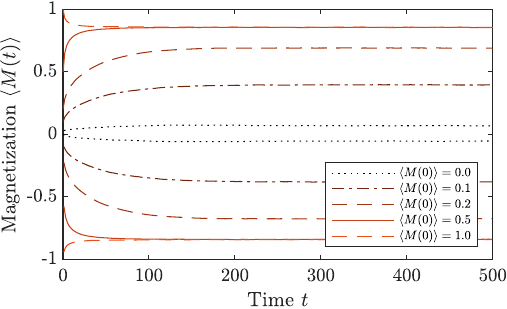}
    \caption{Expectation of magnetization as a function of time $t$ starting from random $X_j$-basis product states with varying mean magnetization (different shades of red and line type); even and odd times are plotted separately, $L = 20$. The data are averaged over $10^4$ samples. All parameters are as in Fig.~\ref{fig:Oscillations}.}
    \label{fig:randomInit}
\end{figure}

\textit{Initial state dependence.---}The results shown in the main text (Fig.~\ref{fig:Oscillations}) are for a system that is initialized in a perfectly magnetized state $\rho(0) = \ket{+^{\otimes N}} \bra{+^{\otimes N}}$. However, a true time crystal should exhibit such oscillations for generic initial states \cite{Keyserlingk2016a}. In an MBL-DTC, the initial state $\rho(0)$ determines the expectation values of l-bit operators $\tau_j^z$, which evolve according to $\tau_j^z(t+1) = -\tau_j^z(t)$. The amplitude of the oscillations of a particular observable $A$ is then determined by these expectation values $\Tr[\rho(0) \tau_j^z]$ combined with the operator-space overlaps of $A$ with $\tau_j^z$ (compound l-bit operators $\tau_{j_1}^z \tau_{j_2}^z \cdots $ should also be included). For typical initial states with low enough entanglement (e.g.~product states), the oscillation amplitude of some appropriately chosen local observable $A$ will indeed be non-zero, and so DTC order can be observed without having to prepare a particular initial state.

In our non-unitary model, there is not an extensive number of emergent conserved quantities; rather the space of stable states has dimension 2. Thus, in the DTC phase (in the thermodynamic limit) it is possible to identify two right eigenoperators of $\mathcal{N}$ --- $\rho_{\rm ss}$ and $\tau$ --- with traces 1, 0, and corresponding eigenvalues 1, $-1$, respectively. Without any fine-tuning, all other eigenoperators of $\mathcal{N}$ will have modulus less than 1, so in the late-time limit ($1 \ll t \ll e^{L/\xi}$), the state of the system will be $\rho(t) = \rho_{\rm ss} + \alpha (-1)^t \tau$, where the coefficient $-1 \leq \alpha \leq 1$ is determined by the initial state. (The coefficient of $\rho_{\rm ss}$ must be unity to ensure the correct trace, and we normalize $\tau$ such that $\Tr[\tau^\dagger \tau] = 1$.) One can then view the extremal density matrices $\rho_{\pm} \coloneqq \rho_{\rm ss} \pm \tau$ as basins of attraction in the space of possible states \cite{Gambetta2019}. An arbitrary initial density matrix can be split into one part that lies in the basin of $\rho_+$, and another in the basin $\rho_-$; the relative weights of these two parts is what determines $\alpha$. Thus, we expect that our system will display robust oscillations for generic initial states, but the mechanism is somewhat different to the MBL-DTC.

To confirm this expectation, we have calculated the time-dependence of the expectation value of magnetization $M = N^{-1} \sum_j X_j$ for systems that are initialized in random product states --- specifically each qubit is in an
eigenstate of $X_j$ with eigenvalue $+1$ ($-1$) with probability $(1 + \langle M(0) \rangle)/2$ ($\langle M(0) \rangle$ is the mean initial magnetization). The results are plotted in Fig.~\ref{fig:randomInit}. When $\langle M(0) \rangle$ is high enough, most initial states being sampled have a sufficiently large net magnetization to lie entirely within one basin of attraction, so $\alpha = 1$. We see that this is the case when $\braket{M(0)}$ is set to $0.5$ and $1$. As $\langle M(0) \rangle$ it becomes increasingly likely for the initial state of a particular sample to lie within the opposite basin, leading to a decrease of the average value of $\alpha$, and in turn oscillations with smaller amplitude. The width of the distribution of initial magnetization decreases with increasing system size, so we expect that the modulation of $\alpha$ with $\langle M(0) \rangle$ will be less severe with increasing $L$ (we have confirmed this with simulations for different $L$). Note that in our simulations, the Ising symmetry is weakly broken, which means that the fine-tuned point where $\alpha = 0$ and the oscillations vanish is not constrained to occur at zero average initial magnetization.\\

\begin{figure}
    \centering
    \includegraphics{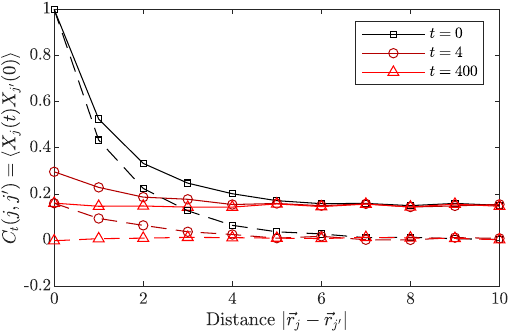}
    \caption{Behaviour of the autocorrelator $C_t(j,j')$ as a function of the distance between qubits, for different times, on a $L \times L$ square lattice with $L = 20$. The coordinates of the qubits are $j = (0,0)$, $j' = (x, 0)$, for various $x$. The solid lines are for a system in the DTC phase (same parameters as Fig.~\ref{fig:Oscillations}), while the dashed lines are for a system where $p_{\rm unit}$ is increased to $0.1$, putting it in the paramagnetic phase. In the limit of large separation $|\vec{r}_{j\vphantom{'}} - \vec{r}_{j'}| \rightarrow \infty$, the autocorrelator approaches a non-zero distance-independent value in the DTC phase, indicating the presence of spatiotemporal order. In the paramagnetic phase, the autocorrelator approaches zero as distance and/or time is increased.}
    \label{fig:autocorrelator}
\end{figure}

\textit{Autocorrelator.---}The autocorrelator is defined as $C_t(j,j') \coloneqq \langle X_j(t) X_{j'}(0) \rangle$, where the expectation value is taken with respect to a steady state of the dynamics. In the DTC phase, this steady state consists of an equal-weight combination of positive- and negative-magnetized oscillating states. We prepare this state by initializing the system in the `paramagnetic' state $\ket{0^{\otimes N}}$ (which has no bias between $+X_j$ and $-X_j$ magnetizations), and evolving for a sufficiently long time such that expectation values of observables become time-independent.  We then perform a projective measurement of $X_{j'}$, giving a random outcome $m = \pm 1$, evolve for a further time $t$, and finally compute the expectation value of $X_j$ and multiply it by $m$. We repeat this, stochastically sampling the measurement outcomes $m$ each time, to obtain an estimate of the autocorrelator. The total number of repetitions per data point of our simulation is $10^5$.

Results for two different sets of parameters are shown in Fig.~\ref{fig:autocorrelator}, for a system of size $L = 20$. The solid lines correspond to the same parameters as in Fig.~\ref{fig:Oscillations}, which belongs to the DTC phase, while for the dashed lines the rate of unitary gates $p_{\rm unit}$ is set to $0.1$, which drives the system into a paramagnetic phase. We see that $C_t(j,j')$ saturates to a non-zero value in the limit of large distance and/or long times in the DTC phase, thus demonstrating the existence of true spatiotemporal order \cite{Khemani2017}. At odd times (not shown), the sign of this value is reversed, which distinguishes the DTC from the ferromagnetic phase. In the paramagnetic phase, correlations decay rapidly with increasing distance and time.\\

\subsection*{Critical Behaviour}

\begin{figure}
    \centering
    \includegraphics[width=246pt]{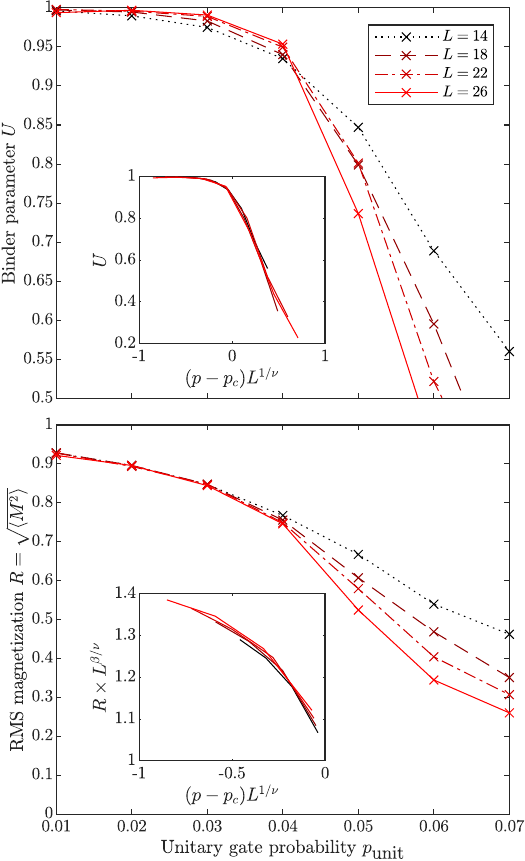}
    \caption{Critical behaviour of the Binder parameter $U = (3 - \braket{M^4}/\braket{M^2}^2)/2$ (top), where $M = N^{-1}\sum_j X_j$ is the sample magnetization, and the RMS magnetization $R = \sqrt{\braket{M^2}}$ (bottom), using the unitary gate probability $p_{\rm unit}$ as a tuning parameter. Other parameters are $p_{\rm NEC} = 0.8$, $p_{\rm flip} = 0.95$; we do not include resets and measurement errors. We estimate the critical value of the unitary gate probability to be $p_{\rm unit} = p_c \approx 0.0427$ using the crossing point of the Binder parameter. Insets: Scaling collapse of the data using the 2D Ising critical exponents $\nu = 1$, $\beta = 1/8$.}
    \label{fig:scaling}
\end{figure}

In this section, we discuss the critical behaviour of our model when tuned close to a phase transition. We will focus on the transition between time-crystalline and paramagnetic phases, using the rate of entangling unitary gates $p_{\rm unit}$ as a tuning parameter. In the following, we set $(p_{\rm flip}, p_{\rm NEC}, p_{\rm reset}, p_{\rm ME}) = (0.95, 0.8, 0, 0)$. The dynamics is therefore symmetric under the up-down symmetry $X_j \rightarrow -X_j$.

Firstly, we consider the instantaneous properties of the steady states to extract static critical properties. To do so, we initialize the system in an up-down-symmetric state (in particular we choose $\ket{0}^{\otimes N}$), and evolve until a steady state is reached. The magnetization $\braket{M}$ (where $M = N^{-1}\sum_j X_j$) must vanish by symmetry, but the even moments $\braket{M^2}$, $\braket{M^4}$ provide information about the symmetry-breaking order. As well as the RMS magnetization $\sqrt{M^2}$, a useful quantity is the Binder parameter $U = (3 - \braket{M^4}/\braket{M^2}^2)/2$ \cite{Binder1981}, which in classical Ising models approaches 1 in the ordered phase, and 0 in the disordered phase. Finite-size scaling can be used to identify the critical point, where the value of the binder parameter becomes system-size independent. When the location of the critical point is known, it is then possible to extract critical exponents for the transition.

\begin{figure}
    \centering
    \includegraphics[width=228pt]{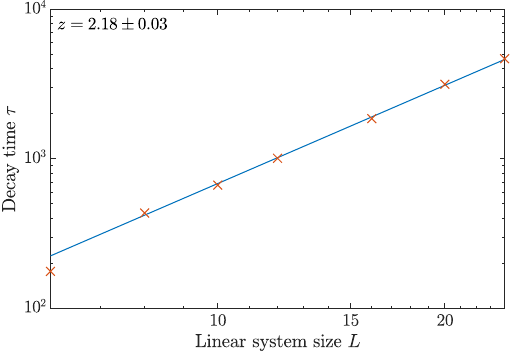}
    \caption{Dependence of the decay timescale $\tau$ [calculated as in Fig.~\ref{fig:Oscillations}(b)] on the linear system size $L$ under dynamics tuned to the critical point, plotted on a log-log scale. Parameters are as in Fig.~\ref{fig:scaling}, with $p_{\rm unit} = p_c \approx 0.0427$. We fit the data to a power law $\tau \propto L^z$, and find a dynamical critical exponent of $z = 2.18 \pm 0.03$. }
    \label{fig:dynamicalexponent}
\end{figure}

Our results are shown in Fig.~\ref{fig:scaling}. We find a critical unitary gate probability of $p_c \approx 0.0427$. Using this value, we find that the data collapses well using the 2D Ising critical exponent $\nu = 1$. The critical behaviour of the Ising symmetry-breaking order parameter $R = \sqrt{M^2}$ is also shown. Again, we see a good collapse of data using the Ising critical exponent $\beta = 1/8$. This indicates that the transition is in the 2D Ising universality class. 

Secondly, we focus on dynamical critical behaviour. In particular, to extract the dynamical critical exponent $z$, we calculate the dependence of the decay timescale of oscillations on the system size $L$, which at criticality should follow $\tau \propto L^z$. This data is shown in Fig.~\ref{fig:dynamicalexponent}, for systems with $p_{\rm unit}$ tuned to the critical value $p_{c} \approx 0.0427$. By fitting a power law to the data, we extract a value of $z = 2.18 \pm 0.03$. This is consistent with the best estimates obtained for critical Glauber dynamics of the 2D classical Ising model $z = 2.1667(5)$ \cite{Nightingale2000}. We have confirmed that this value remains unchanged along the phase boundary by performing identical simulations at a critical point where we set $p_{\rm NEC} = 0.7$, giving a new critical unitary gate probability $p_c' \approx 0.0266$; the power law fit there gives a consistent value of $z = 2.15 \pm 0.03$.

The collection of critical exponents we find (both static and dynamical) are the same as those that would be obtained for classical stochastic dynamics at a phase transition in the 2D Ising universality class. Two comments on this finding are required. Firstly, it is interesting that the critical behaviour of this strongly non-equilibrium model, which does not respect detailed balance, coincides with that of an equilibrium statistical mechanics model. This fits with arguments that (at least for classical systems) the critical exponents for systems with non-conserved order parameter match that would be found in an equilibrium model \cite{Tauber2002}. Note, however, that we expect that certain critical correlators will be sensitive to the lack of microreversibility in our system, as occurs in, e.g.~Ref.~\cite{Gambetta2019a}. Secondly, the appearance of 2D Ising exponents is notably different from what one would expect in the vicinity of a quantum phase transition under unitary dynamics. For instance, in the models discussed in Ref.~\cite{Isakov2011} (under sufficiently generic perturbations), the static critical exponents are those of the 3D Ising universality class, reflecting the effective extra dimension given to the theory under the quantum-classical correspondence. Here, the dynamics is intrinsically non-unitary, and so the state of the system is generically mixed. Thus, the critical theory is closer in nature to a finite-temperature phase transition, rather than being described by a conformal quantum critical point as in \cite{Isakov2011}. This explains why we find 2D, rather than 3D, Ising critical exponents.

\subsection*{Numerical simulations beyond Clifford circuits}

\begin{figure}[!t]
    \includegraphics[width=\linewidth]{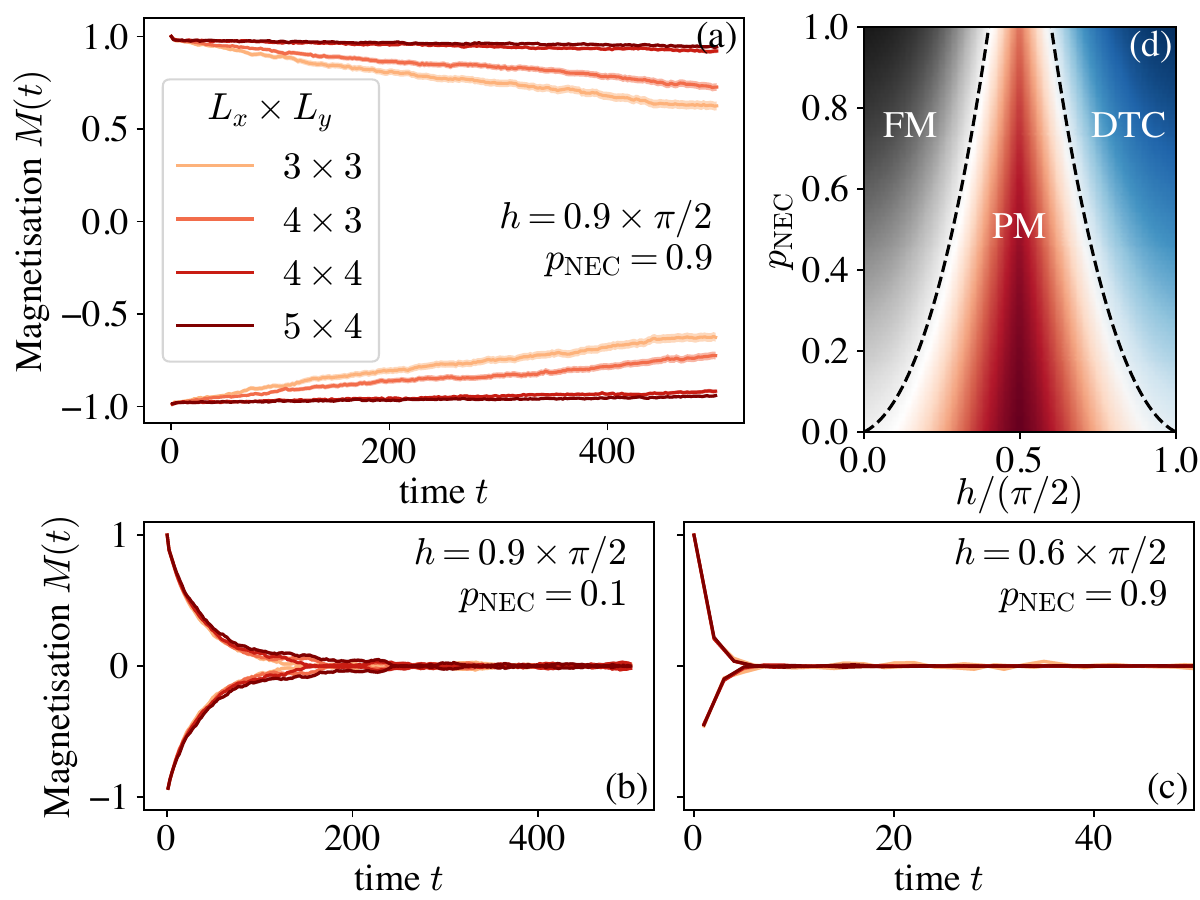}
    \caption{Numerical results for the model described via Eqs.~\eqref{eq:nonclifford} and \eqref{eq:Mtnoncliff}. Panels (a)-(c) show the magnetisation as a function of time for three different values of $(h,p_\mathrm{NEC})$ where the different shades denote different system sizes and the two branches denote odd and even stroboscopic times. Panel (d) shows a conjectural phase diagram of the dynamics in the two-parameter space of $h$ and $p_\mathrm{NEC}$ with the gray, red, and blue regions corresponding to ferromagnetic, paramagnetic, and time-crystalline spatiotemporal order respectively.}
    \label{fig:nonclifford}
\end{figure}

In this section, we provide proof-of-principle results showing the presence of spatiotemporal order in a setting less restrictive than the Clifford dynamics described until now. The system again comprises of qubits on a rectangular lattice that undergo dynamics described by 
\begin{equation}
    \ket{\psi(t+1)} = \frac{\mathcal{M}_tU^{(X)} U_t^{(Z)}\ket{\psi(t)}}{||\mathcal{M}_tU^{(X)} U_t^{(Z)}\ket{\psi(t)}||}\,,
    \label{eq:nonclifford}
\end{equation}
where $U^{(Z)}_t = \exp[\sum_j h_{j,t}Z_j]$ with $h_{j,t}\in[h-\delta h,h+\delta h]$ encodes the noisy, disordered (random in space and time) spin-flips, $U^{(X)}=\sum_{\braket{i,j}}J_{ij}X_iX_j$ with $J_{ij}\in[J-\delta J,J+\delta J]$ encodes the interaction between nearest-neighbour qubits on the lattice, and $\mathcal{M}_t$ denotes the measurement and feedback step. The final step, $\mathcal{M}_t=\prod_j\mathcal{M}_{j,t}$, is implemented as 
\begin{equation}
    \mathcal{M}_{j,t}=
    \begin{cases}
    \mathbb{I} & \mathrm{with~probability~~} 1-p_\mathrm{NEC}\\
    U_{w_{\hat{n}},w_{\hat{e}}}\prod_{j,\hat{n}}^{w_{\hat{n}}}\prod_{j,\hat{e}}^{w_{\hat{e}}} &\mathrm{with~probability~~}p_\mathrm{NEC}p_{j,t}^{w_{\hat{n}},w_{\hat{e}}}
    \end{cases}\,,
        \label{eq:Mtnoncliff}
\end{equation}
where $p_\mathrm{NEC}$ is the probability of measurement and $p_{j,t}^{w_{\hat{n}},w_{\hat{e}}}=\braket{\psi(t)|\prod_{j,\hat{n}}^{w_{\hat{n}}}\prod_{j,\hat{e}}^{w_{\hat{e}}}|\psi(t)}$ is the Born rule probability. We fix $J=1$ and $\delta J=0.2=\delta h$ and study the problem in a two-dimensional parameter space spanned by $h$ and $p_\mathrm{NEC}$. Loosely speaking, $h=0$ and $h=\pi/2$ can be considered analogous to the $p_\mathrm{flip}=0$ and $p_\mathrm{flip}=1$ limits of the Clifford circuit discussed in the main text.

Within this setting, let us now turn towards the results. For simplicity we start from a fully polarised state $\ket{\psi(0)}=\ket{+^{\otimes N}}$ and track the magnetisation $M(t)=N^{-1}\sum_i \braket{X_i(t)}$. We expect a robust phase with a time-crystalline spatiotemporal order in the vicinity of $h\sim\pi/2$ and $p_\mathrm{NEC}\lesssim 1$. This is indeed bourne out in Fig.~\ref{fig:nonclifford}(a) where the lifetime of the time-crystalline order clearly increases with increasing system size for $h=0.9\times\pi/2$ and $p_\mathrm{NEC}=0.9$. On decreasing $p_\mathrm{NEC}$, we expect that below a threshold value the corrective feedback becomes weak enough that the inherently noisy and hence thermalising unitary part of the dynamics in Eq.~\eqref{eq:nonclifford} melts the spatiotemporal order. This is indeed what we see in Fig.~\ref{fig:nonclifford}(b) for $p_\mathrm{NEC}=0.1$. The spatiotemporal order can also be destroyed by tuning $h$ well away from $\pi/2$ such that the spin-flips due to $U^{(Z)}_t$ are effectively weak enough to not induce the temporal order [see Fig.~\ref{fig:nonclifford}(c)]. By analogous considerations in the vicinity of $h\sim 0$, we expect the same phenomenology but for a phase with non-trivial spatial order but no non-trivial temporal order. We do indeed find such a measurement-feedback stabilised ferromagnetic phase but do not show the results for brevity. All of the above can be summarised via the schematic phase diagram shown in Fig.~\ref{fig:nonclifford}(d).

\end{twocolumngrid}


\end{document}